\documentclass[aps,showpacs,superscriptaddress,twocolumn]{revtex4}
\usepackage{epsfig}

\def\be{\begin{equation}}
 \def\ee{\end{equation}}
\def\bea{\begin{eqnarray}}
 \def\eea{\end{eqnarray}}
\def\half{\frac{1}{2}}

\def\PRB{Phys. Rev. B~}

\def\PRL{Phys. Rev. Lett.~}
\def\PR{Phys. Rev}
\def\LCO{La$_2$CuO$_4$}
\def\LSCO{La$_{2-x}$Sr$_x$CuO$_4$}
\begin{document}

\title{ A Magnetic Model of the
Tetragonal-Orthorhombic Transition in  the Cuprates}

\author{Fang Chen, Jiangping Hu}
\affiliation{ Department of Physics, Purdue University, West
Lafayette, IN 47907 }
\author{Steven Kivelson}
\affiliation{Department of Physics, McCullough Building, Stanford
University, Stanford CA 94305-4045}
\author{Stuart Brown}
\affiliation{Department of Physics \& Astronomy, University of
California at Los Angeles, Los Angeles CA 90095}

\begin{abstract}
It is shown that a quasi two dimensional (layered) Heisenberg
antiferromagnet with fully frustrated interplane couplings ({\it
e.g.} on a body-centered tetragonal lattice) generically exhibits
two thermal phase transitions with lowering temperature -- an
upper transition at $T_{TO}$ (``order from disorder without
order'') in which the lattice point-group symmetry is
spontaneously broken, and a lower N\'{e}el transition at $T_{N}$
at which spin-rotation symmetry is broken. Although this is the
same sequence of transitions observed in La$_2$CuO$_4$,  in the
Heisenberg model (without additional lattice degrees of freedom)
$(T_{TO}-T_N) /T_N$ is much smaller than is observed. The model
may apply to the bilayer cuprate La$_2$CaCuO$_6$, in which the
transitions are nearly coincident.
\end{abstract}
\maketitle

\section{Introduction}

In addressing the physics of high temperature superconductivity
(HTC), an important, but often overlooked, issue is the relation
between  the lattice structure and the electronic physics. In the
highly studied ``214" family of HTC superconductors, there is a
structural phase transition from a high temperature tetragonal
 (HTT) to a low temperature orthorhombic (LTO) phase
with a transition temperature, $T_{TO}(x)$ which drops as a
function of doped hole concentration, $x$, from $T_{TO}(0)\approx
500$K in undoped {\LCO}, to $T_{TO}(0.22)=0$ in {\LSCO} with $x\ge
0.22$ \cite{Paul1987,Fleming1987,Braden1994,Gilardi2000}. We were
initially motivated to examine the problem of interlayer magnetic
coupling because, in the N\'{e}el state, the coupling between
near-neighbor planes is frustrated in both the T-phase of the 214
compounds ({\it e.g.}, LSCO and Sr$_2$CuO$_2$Cl$_2$
\cite{Greven1994}), and the T'-phase ({\it e.g.} Pr$_2$CuO$_4$ or
Nd$_2$CuO$_4$) \cite{Tokura}) due to the location of the Cu$^{2+}$
ions on a body-centered tetragonal lattice. The HTT-LTO transition
of the T-phase removes the frustration, and always occurs well
above any magnetic ordering temperature, $T_{N}$. On the other
hand, no lattice distortion has been identified in the T' systems,
even though the transition temperature to the magnetically ordered
N\'{e}el state is similar for compounds with the two structures
\cite{Matsuda1990,Keimer1992}.

The question we address is how a purely electronic system, with no
lattice coupling, could spontaneously break the point group
symmetry which results in the perfect cancellation of the magnetic
coupling between the sublattices. This approach is contrary to
conventional wisdom, which holds that the HTT-LTO transition in
LCO is driven by a lattice parameter mismatch between the CuO$_2$
planes and the interstitial layer La-O. There are, however,
reasons to question whether this explanation is complete:  {\bf
1)} Doping with Ba ions could be expected to relieve the stress
more efficiently than doping with Sr, because of its larger ionic
radius. It doesn't seem to do that: the T$_{TO}\to 0$ at roughly
x=0.20 in both cases. {\bf 2)} The in-plane magnetic
susceptibility is found to be spectacularly anisotropic
\cite{Lavrov2001,Lavrov2002} below T$_{TO}$ ($\chi_{bb}$ is
between two and three times larger than $\chi_{aa}$, where $a$ and
$b$ refer to the orthorhombic a and b axes); were the electronic
anisotropy simply a response to the symmetry breaking lattice
distortion, then especially at temperatures above $T_{N}$ but
below $T_{TO}$, one would expect the anisotropy of any electronic
response to be small in proportion to the magnitude of the
orthorhombic distortion, which is around 4\% in these materials.

We consider undoped {\LCO} where, due to the strong repulsions
between electrons, the low energy electronic physics is well
approximated by a quantum Heisenberg antiferromagnet with a single
spin 1/2 on each Cu site. We define a continuum field theory (a
$O(N)$ non-linear sigma model with $N=3$) in terms of the
corresponding long-wave-length excitations, and solve it in the
large $N$ limit ({\it i.e.} in the self-consistent phonon
approximation).  We show that the spin model undergoes a similar
sequence of two phase transitions upon cooling as does {\LCO} -- a
high temperature T-O transition and a lower temperature
transition, $T_{N}< T_{TO}$, at which spin-rotation invariance is
spontaneously broken resulting in a Neel state. (This corresponds
to the  schematic phase diagram in Fig. 2c.) The T-O transition in
this model is an Ising transition and is an example of ``order by
disorder without order", as we explain below. The magnetic order
is collinear, consistent with what is determined for {\LCO}.
However, the difference between the two transition temperatures is
very small, $(T_{TO} - T_N)/T_N \ll 1$, while what is observed in
{\LCO} is a substantial difference, $(T_{TO} - T_N)/T_N \sim 1$;
thus, it is clear that the lattice degrees of freedom play a very
significant (possibly dominant) role in the stabilization of the
LTO phase in these materials. On the other hand, $T_{TO}$ and
$T_N$ are very close in the bilayer cuprate La$_2$CaCuO$_6$
\cite{Ulrich2002,Hucker2005}, and the model may apply in this
case.


\section{The microscopic model}
We start by defining a model of $M$ coupled layers, each of which
consists of a square lattice Heisenberg antiferromagnet, with the
sites on one layer situated above the plaquette center of the
layer below: \bea H&&=\sum_{n=1}^M\sum_{\vec R}\sum_{\vec\delta_0}
\frac J 2 \ {\bf S}_{n,\vec R}\cdot {\bf S}_{n,\vec R+\vec
\delta_0}
\\ &&+\sum_{n=1}^{M-1} \sum_{\vec R}\sum_{\vec \delta_1}J_n^\prime\ {\bf S}_{n,\vec R}\cdot {\bf S}_{n+1,\vec R+\vec \delta_1}
\nonumber \label{H} \eea where  ${\bf S}_{n,\vec R}$ is the spin
at site $\vec R$  in plane $n$ and the sums over $\vec \delta_0$
and $\vec\delta_1$ run, respectively, over pairs nearest-neighbor
sites within a plane and in neighboring planes: $\vec\delta_0=\pm
\hat x$, $\pm \hat y$ (in units in which the in-plane lattice
constant $=1$) and $\vec\delta_1=\pm  (1/2)[\hat x+\hat y]$, $\pm
(1/2)[\hat x-\hat y]$.  Below, we will consider the  bilayer
($M=1$)  and the   body-centered tetragonal ($M\to\infty$)
versions of this problem.  We will also consider cases in which
the spacing between the layers are not all the same so that
$J_n^\prime$  is $n$ dependent, but  we will always assume $J \gg
J^\prime>0$.

In a classically Neel ordered state, ${\bf S}_{n,\vec R} = {\bf
M}_n\exp\{i\vec Q\cdot\vec R\}$ with $\vec Q=\langle
\pi,\pi\rangle$, there is no coupling between the staggered order
parameters, ${\bf M}_1$ and ${\bf M}_2$, in the two planes.
However, high energy short wave-length quantum fluctuations induce
an additional effective interaction between the spins at low
energy, $H\to H+H_{ind}$: \be H_{ind}=-\sum_{n=1}^{M-1}\sum_{\vec
R } K_n\left[\Phi_{n,\vec R}\right]^2 \label{Hind} \ee where
$\Phi_{n,\vec R}\equiv\sum_{\vec \delta_1}[ e^{i\vec
Q\cdot\vec\delta_1}{\bf S}_{n,\vec R}\cdot{\bf S}_{n+1,\vec
R+\vec\delta_1 }]$ is a composite order parameter field and
\footnote{Were we to  derive $K_n$ from a spin-wave expansion for
large spin, $S$, we would find that $K_n \sim
S^{-3}(J_n^\prime)^2/J$. However, since in many of the physical
realizations of this model, $S$ is 1/2 or 1, it is better to think
of generating $K_n$ in the first stage of a renormalization group
proceedure in which short-wave-length, high energy modes are
integrated out.} $K_n \sim (J_n^\prime)^2/J$.    For $K_n >0$, the
staggered moments on neighboring planes tend to be colinear to
maximize the expectation value of $\Phi$.

Rotation by $\pi/2$ about an axis through any lattice site $\vec
R$ on plane $n$ is a symmetry of the Hamiltonian.  Such a rotation
transforms $\phi_{n,\vec R} \to -\Phi_{n,\vec R}$, so any state
with $\langle \Phi\rangle \ne 0$ is a state of broken symmetry.
For instance, the state pictured in Fig. 1 has $\langle
\Phi_{1,\vec R}\rangle < 0$, but upon rotation by $\pi/2$ about a
site in the lower plane, one obtains the state in which all the
spins in the upper layer are reversed, and hence $\langle
\Phi_{1,\vec R}\rangle$ changes sign. More generally, a non-zero
$\langle \Phi_{1,\vec R}\rangle$  occurs only in a state in which
the discrete C$_4$ rotational symmetry of the lattice is broken
down to C$_2$, {\it i.e.} in an ``Ising Nematic'' state, or
equivalently an electronically orthorhombic state.

\section{Order from disorder}
{\it In two dimensions:} A mechanism is called ``order by
disorder" when it is driven by fluctuation induced terms that are
absent at the mean-field level
\cite{Shender1982,Chandra1990,Henley1989}. A classic example is
the bilayer of Heisenberg antiferromagnet shown in Fig. 1(the case
$M=1$).  Here, in addition to the usual ground-state degeneracy
which is  implied by the spontaneous breaking of spin-rotation
symmetry, at mean-field level there is an added (continuous)
degeneracy  due to the frustrated character of the intersystem
coupling;  the orientation of the staggered magnetization on the
upper layer can be rotated by an arbitrary angle relative to that
on the lower layer at no cost in energy. However, quantum
fluctuations (Eq. \ref{Hind}) lift this degeneracy and produce a
preferred relative orientation such that the staggered
magnetization on the two planes is colinear.   As we have seen,
this state also breaks the discrete rotational symmetry of the
lattice.

The fluctuations that lead to order by disorder are predominantly
short-wavelength, high energy quantum fluctuations, and hence they
produce the same tendency for the spins on neighboring planes to
be colinear, even when there is no long-range spin order.  For
instance, since the bilayer problem is a 2D problem, at any finite
temperature, $T$, spin rotation symmetry cannot be spontaneously
broken.  However, the Ising Nematic order can (and does) survive
to finite $T$ \cite{Cap,Web,Biskup2004}. Thus, although originally
conceived as being a consequence of small fluctuations about a
magnetically ordered state, the tendency of these fluctuations to
produce this order persists even when long-wavelength fluctuations
have destroyed the antiferromagnetic order, itself, leaving
$\xi_{S}(T) < \infty$ -hence \cite{Biskup2004} ``ordering by
disorder without order.''

We can estimate the transition temperature as being the point at
which the induced interaction between correlated blocks of spins
is of order $T$ (in units in which $k_B=1$) \be K\xi_S^2 \sim
T_{TO}^{(bi)} \label{twodim}
 \ee Since $\xi_S\sim \exp[2A\pi J/T]$, the ordering
temperature, $T_{TO}^{(bi)} \sim J/\log[J/K]$ is significantly
smaller than $J$ only if $K$ is astronomically small.  (Here
$A\approx 1/4$ for spin 1/2 due to quantum corrections.)  The
resulting phase diagram is shown
  in  the schematic phase diagram (a) in Fig. \ref{fig:layer}.

{\it In three dimensions:}  The magnetic Hamiltonian on the HTT
lattice is the three dimensional extension of the same model,
where  $M\to \infty$ and $J^\prime_n$ and $K_n$ are indenpendent
of the layer index, $n$. However, to place the problem in a
broader context, in Fig. 2 we have considered a somewhat more
general  version of this Hamiltonian in which $J \gg
J_{2n+1}^\prime \equiv J^\prime \ge J_{2n}^\prime\equiv
J^{\prime\prime}>0$ and $K_n =K(J_n/J^\prime)^2$.  For
$J^{\prime\prime}/J^\prime =1$, this corresponds to the LTT
structure of {\LCO}, while for $J^{\prime\prime}/J^\prime =0$ this
corresponds to an array of decoupled bilayers.

The behavior of the phase boundaries for $1 \gg
J^{\prime\prime}/J^\prime >0$ can be readily deduced from the
single bilayer results by scaling.  The Neel temperature rises
steeply from zero as $T_{N} \sim 2A\pi
J/\log[J/J^{\prime\prime}]$. The T-O transition rises from its
finite bilayer value as $T_{TO}-T_{TO}^{(bi)} \sim J
(J^{\prime\prime}/J)^{2/\gamma}$, where $\gamma = 7/4$ is the
susceptibility exponent of the 2D Ising model.

To obtain a  general solution valid for $J^\prime \sim
J^{\prime\prime}$, we can no longer perturb about the isolated
bilayer, and so must rely on other,  approximate methods.

\section{Continuum O(N) model }
 We start
with the following continuum Hamiltonian, \begin{widetext}
\bea
 H_{c}=\int\int dxdy\sum_{n=1}^{M}
[\half(\rho_s(|\nabla {\bf \phi}_n |^2+|\nabla{\bf \psi}_n |^2) -g_1
({\bf \phi}_n  \cdot {\bf \psi}_n )^2+\frac{\eta_1}{2}{\bf \phi}_n
\partial_x\partial_y {\bf \psi}_n   -g_2({\bf \phi}_{n} \cdot{\bf
\psi}_{n-1})^2+\frac{\eta_2}{2}{\bf \phi}_n
\partial_x\partial_y {\bf \psi}_{n-1}  ] \label{Hc}
 \eea
 \end{widetext}
where ${\bf  \psi}_n(\vec r)$ and ${\bf \phi}_n(\vec r)$ are the
local staggered magnetization order parameters in  even and odd
layers and $|{\bf \phi}_n(x,y)|^2=|{\bf \psi}_n(x,y)|^2=1$. Here
the spin stiffness $\rho_s \propto J$, $g_1\propto \frac{J^{\prime
2}}{J}$, $g_2\propto\frac{J^{\prime\prime 2}}{J}$, $\eta_1 \propto
J^\prime$ and $\eta_2 \propto J^{\prime\prime} $. (We have
considered multiple coupled bilayers, but by making $g$ and $\eta$
dependent on layer index, the more general model can be studied as
well.) We also take the order parameters to be $N$ component
vectors, generalizing from the physical $N=3$, since in the large
$N$ limit (which we expect to be qualitatively correct for $N=3$)
the phase diagram of above model can be obtained as follows.  The
partition function is given by \bea Z =\int D\lambda^1 D\lambda^2
D\sigma D\phi D\psi exp(-\int\int dxdy\sum_n L_n) \eea where the
Lagrangian is given by
\begin{widetext} \bea
 \frac{L_n}{N}= & &\lambda^1_n(|\vec
\psi_n|^2-1)+\lambda^2_n(|\vec \phi_n|^2-1)+ 2[\sigma_{1n}(\phi_n
\cdot \psi_n )+\sigma_{2n}(\phi_n \cdot \psi_{n-1})]
+\frac{T\sigma_{1n}^2}{g_1}+\frac{T\sigma_{2n}^2}{g_2}+\frac{1}{2T}[\rho_s(|\nabla
\vec \phi_n |^2+|\nabla \vec \psi_n |^2)\nonumber
\\
& &+ \eta_1{\bf \phi}_n \partial_x\partial_y {\bf \psi}_n + \eta_2
{\bf \phi}_n
\partial_x\partial_y {\bf \psi}_{n-1} ] \eea
\end{widetext}
where  $\lambda^i_n(\vec r),i=1,2$ are the Lagrangian multiplies for
$\vec\psi_n(\vec r)$ and $\vec \phi_n(\vec r)$ respectively and
$\sigma_{in},i=1,2$ are the Hubbard-Stratonovich field.  The saddle
point of above Lagrangian  is determined  the following
self-consistent equations by taking  $\sigma_{in}(r)=\sigma_i,
i=1,2$ and $\lambda_n(r)=\lambda$ (subject to the stability
condition $2\lambda\ge |\sigma_1+\sigma_2|$): \bea
\sigma_i&=&\frac{-g_i}{(2\pi)^3} \int d^2\vec k dk_z (cos k_z
G_{00} -(-1)^i sin k_z G_{01})\label{self}\\
1&=&m^2+\frac{T}{(2\pi)^3}  \int d^2\vec k dk_z G_{00} \eea where we
have taken $\lambda^i_n(\vec r)=\lambda, i=1,2$, $m$ is the
staggered moment which is non-zero only in the magnetically ordered
phase, the integral for $\vec k$ has  an ultra-violet cuttoff
$\Lambda$, and
 \bea G(k_x,k_y,k_z)=\left( \begin{array}{cc}
  A(\vec k)
  &
B(\vec k)\\
 -B(\vec k) & A(\vec k)
\end{array}\right)^{-1},
\eea where $A(\vec k)=\lambda+\frac{\rho_s k^2}{k_BT}+\frac{cos
k_z}{2}(\sigma^{+}+ \eta^+k_xk_y)$ and $B(\vec k)= i\frac{sin
k_z}{2}(\sigma^{-}+ \eta^-k_xk_y)$ with
$\sigma^{\pm}=\sigma_1\pm\sigma_2$ and $\eta^{\pm}=\eta_1\pm\eta_2$
.

First, let us look the uniform case $ g_1=g_2=g, \eta_1=\eta_2=\eta$
and $\sigma_1=\sigma_2=\sigma$. If $\eta =0$, the above self
consistent equation can be solved exactly. There is a first order
transition at $T_c$ given by $\frac{g}{4\pi\rho_s}=
\frac{2\rho_s\Lambda}{T_c}e^{-4\pi\rho_s/T_c}$. For $T
> T_c$, these equations have only one solution: $m=\sigma=0, \lambda =
\frac{\rho_s\Lambda}{T}e^{-4\pi\rho_s/T}$.  At $T=T_c$ , there is
a family of solutions with $m=0$ and
$\lambda+\sqrt{\lambda^2-\sigma^2}=\frac{g}{4\pi\rho_s}$.  For
$T<T_c$, the minimal solution has $\sigma=\lambda=g/4\pi\rho_s$
and non-zero magnetization, \be m^2 = \frac {(T_c-T)} {T_c} -
\frac {T_cT} {4\pi\rho_s} \log[T/T_c]. \ee  The transition at
$T_c$ is a peculiar:  The orthorhombic order parameter jumps to a
finite expectation value, $\Phi \propto \sigma=g/4\pi\rho_s$, just
below $T_c$, making it a first order transition.  At the same
time, the magnetization grows continuously from zero as $T$ drops
below $T_c$. (Unsurprisingly, the coincidence of the two
transitions is lifted, as we shall see now, by non-zero $\eta$.)
Of course, because $\lambda=\sigma$, there is a gapless
(Goldstone) mode for all $T\le T_c$.

When $\eta\neq 0$, there are two phase transitions. The first one
is a second order Ising-type transition at $T_{TO}$  and the
second one is the Neel transition at $T_N$.  $T_{TO}>T_N$ for any
nonzero value of $\eta$. The Ising transition is marked at
$T_{TO}$ where $\sigma(T_{TO})=0$ and
$\frac{d\sigma}{dT}|_{T=T_{TO}}\neq 0$ and the Neel transition
happens at $T_N$ where $\lambda=\sigma$. From these conditions,
equations which determine $T_{TO}$ and $T_N$ can be derived. While
it can be proved that the solutions of the equations exist and
satisfy $T_{TO}>T_N$ when $\eta \neq 0$, it would be tedious to
present the proof.

However, when $\eta$ is small, we can solve the critical
temperatures up to the second order of $\eta$. Expanding the self
consistent equations up to the second order of $\eta$ and defining $
  \alpha=\frac{g}{4\pi\rho_s},
\beta=\frac{\eta}{\rho_s}$ and
$\gamma(T)=\frac{exp(\frac{4\pi\rho_s}{T(1+\frac{\beta^2}{16})})}{\frac{2\rho_s\Lambda}{T}}$,
the two critical temperatures are determined by \bea
\frac{\gamma(T_{TO})^{-1}}{\alpha}=1+\frac{3\beta^2}{16},
\frac{\gamma(T_N)^{-1}}{\alpha}=1+\frac{\beta^2}{6}.  \eea From
these equations, it is transparent that $T_{TO}>T_N$. Up to the
second order in $\beta$, the difference between $T_{TO}$ and $T_N$
is given by \bea
    \delta T = T_{TO} -T_N = \frac{T_N^2\beta^2}{48(4\pi\rho_s-T_N)}
    \label{result}
    \eea

While this result shows that the fluctuations will induce the
Ising order transition before the temperature reaches the Neel
transition, 

As $\eta$ is expected to be small, it follows that for the uniform
case, so is $(T_{TO}-T_N)/T_N$. For nonuniform coupling parameters
$g_1\neq g_2$ and $\eta_1\neq\eta_2$, there are also two phase
transitions and the difference between two transition temperatures
can be large. In the large $N$ limit, the Ising transition critical
temperature is mainly determined by $g_1$ and $\eta_1$ while the
Neel transition critical temperature is mainly determined by $g_2$
and $\eta_2$. To show the difference between $T_{TO}$ and $T_N$ in
this dimerized model, we take $J=0.12ev$, $\eta_1/\rho_s=0.1$,
$\rho_s=JS^2$, and $g_i=0.26JS\Lambda\frac{\eta_i^2}{\rho_s^2}$. We
choose the cutoff $\Lambda=100$ and $S=1/2$.  By varying $\eta_2$,
we find that the transition temperature $T_{TO}\simeq 470k$, which
is almost independent of $\eta_2$.  In fig.\ref{fig3}, we plot
$\frac{T_{TO}-T_{N}}{T_{N}}$ as a function of $ln\frac{g_1}{g_2}$.
It fits to a very good linear dependence.  So far, there is no
experimental evidence  supporting  layer dimerization along c-axis
in cuprates.  The quantitative results here can not be directly
linked to LCO. However, it will be interesting to see whether the
results can be applied in other systems.

If just for a test of the accuracy of the large $N$ method used
here, and as a comparison to the result in two dimension, it is
interesting to apply the same method to the bilayer problem. The
continuum Hamiltonian for this case is $H_c$ in Eq. \ref{Hc} with
$M=1$. The corresponding saddle-point equations are obtained from
Eqs. \ref{self} by replacing the integral over $k_z$ with a sum
over $k_z=0$, $\pi$.  These equations are solved with $m=0$ at all
non-zero temperatures. There is, however, a non-zero critical
temperature for the T-O transition, $T_{TO}$, such that $\sigma=0$
for $T \ge T_{TO}$, and $\lambda = (\rho_s\Lambda/T)\exp[-4\pi
A_{\eta}\rho_s/T]$ where $A_{\eta}=\sqrt{1-(\eta T/2\rho_s)^2}$.
Taking this expression for $\lambda(T)$, the critical temperature
is the solution of the implicit equation $4\pi
A_{\eta}\rho_s\lambda=g$, from which it follows that $T_{TO} \sim
4\pi \rho_s/\log[g/\rho_s\Lambda]$. As $T$ is decreased below
$T_{TO}$, $\sigma$ rises continuously from 0.

\section{Discussion}

In $La_{2-x}Sr_xCuO_4$, the HTT to LTO transition occurs at
roughly twice $T_N$, and so is inconsistent with our findings.
Probably, this signifies a significant role of the lattice degrees
of freedom in driving the transition. However,  this does not
necessarily imply that the electronic considerations found here
are entirely unimportant - in particular, the order one anisotropy
in the magnetic susceptibility\cite{Lavrov2001} in the temperature
range between $T_N$ and $T_{TO}$ argues in favor of a significant
electronic contribution to the transition. A more compelling case
for an electronically driven transition to an orthorhombic phase
can be made for the two-layer compound La$_2$CaCu$_2$O$_6$, where
the magnetic and structural transitions are observed and they are
close together in temperature \cite{Ulrich2002,Hucker2005}.

















It is a very general feature of the fluctuation induced coupling
between frustrated planes that $K>0$, and hence that the induced
order is collinear.  However, a finite concentration of vacancies
can induce the opposite sign of interplane coupling which can
therefore lead to non-collinear ordering of magnetic moments in
different layers\cite{Henley1989}. When a vacancy is created in
one plane, the LTO order near the vacancy is destroyed. The effect
of vacancies on $T_{TO}$ is illustrated in Fig. \ref{van}.   At
mean-field level, the net field applied on any spin in one plane
due to the spins in the plane below is zero;  however, where there
is  a vacancy in the lower plane, all the neighboring spins in the
next plane feel a net field which is $J^\prime$ times minus the
spin that was removed to make the vacancy.  As is well known, an
applied field in an antiferromagnet favors the ordered moments in
the plan perpendicular to the field, which permits  a net canting
of the spins in the direction of the applied field -- this is the
driving force for the usual spin-flop transition. Thus, there is a
negative contribution to the average coupling, $ K_{vac} \sim -
n_{vac}J'^2/J$ which is proportional to the vacancy concentration,
$n_{vac}$. We note the AF systems $R_2$CuO$_4$ ($R$= Nd,Pr,Gd,Eu)
\cite{Sumarlin,Braden,Ska,Chattopadhyay1994} forming in the T'
structure exhibit the combination of no detectable orthorhombic
distortion {\it and} non-collinear ground states. Possibly, the
difference of the T' compounds from La$_2$CuO$_4$ is the moment
associated with the rare earth atoms( for Eu atoms, due to the
relatively small energy gap between the first excited state and
the ground state, they carry significant magnetic moments at 300K
although the ground state  is nonmagnetic); including them leads
to a magnetic Hamiltonian different from Eq. (1).

{\it Acknowledgments:} We thank R.J. Birgeneau, M. Greven, B.
Keimer, E. Shender, and J. Tranquada for helpful discussions. F.
Chen and J.P Hu gratefully acknowledge support by a Purdue
research grant. This work was supported in part by the National
Science Foundation under grant nos. DMR-0421960 (SAK) and
DMR-0520552 (SEB).

\begin{figure*}
\includegraphics[width=5cm, height=4cm]{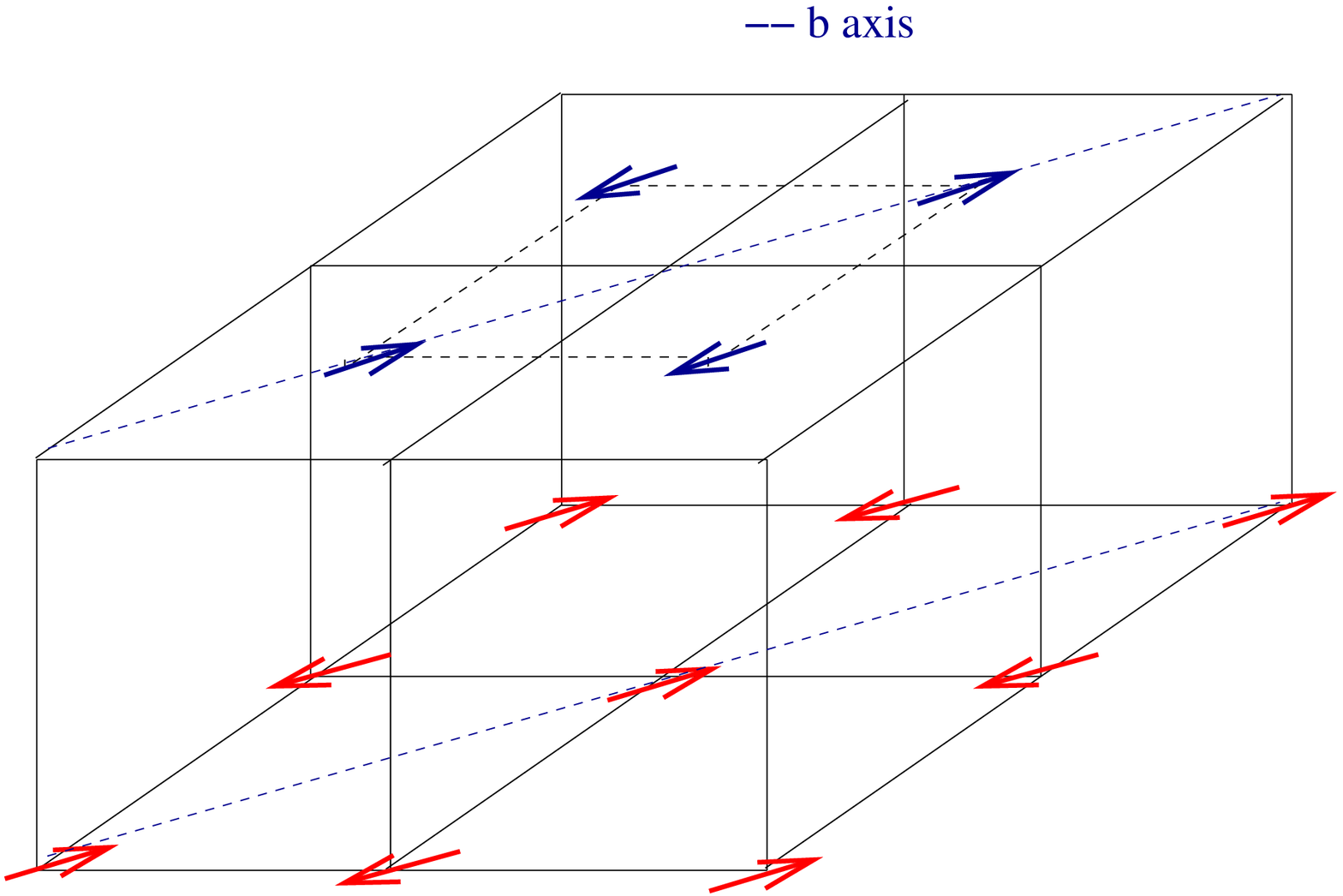}
\caption{\label{fig1}  A sketch for the bilayer Heisenberg model
in square lattice.}
\end{figure*}

\begin{figure*}
\includegraphics[width=6cm, height=5cm]{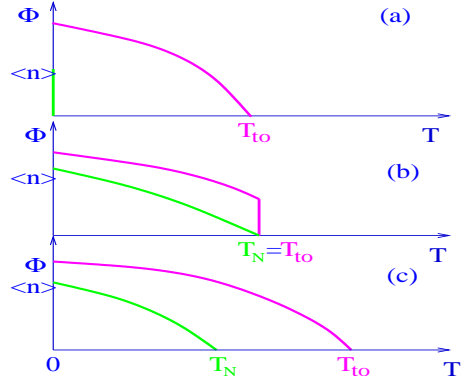}
\caption{\label{fig:layer} a) The phase diagram for a bilayer
system: $T_N=0$ $T_{to}\neq0$. b)The phase diagram for uniform case
with  $\eta=0$. The transition is first order and $T_{to}=T_N$ c)
The general phase transition. There are two second order
transitions: $T_{to}>T_N$. }
\end{figure*}

\begin{figure*}
\includegraphics[width=6cm, height=5cm]{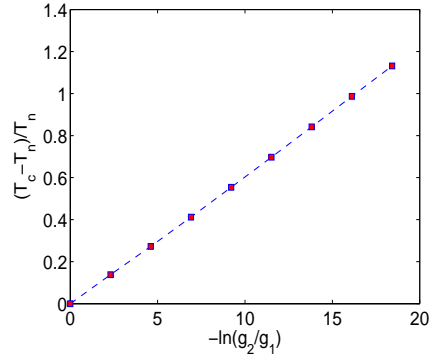}
\caption{\label{fig3}  The difference between two critical
temperatures $\frac{T_{to}-T_{N}}{T_N}$ as a function of
$ln(g_1/g_2)$.}
\end{figure*}

\begin{figure*}
\includegraphics[width=5cm, height=4cm]{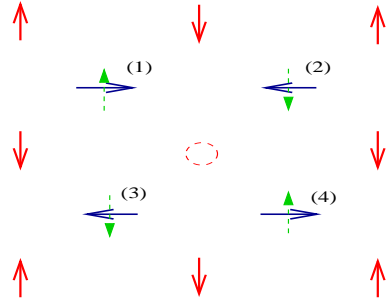}
\caption{\label{van} A sketch of spin configurations in presence of
one vacancy.  When a vacancy is created in one plane, spins in the
nearest neighbor plane favor non-collinear alignment. If alignment
is collinear, which is illustrated by dashed arrows, the third and
fourth spins are not happy energetically.}
\end{figure*}

\end{document}